# Odd J states of isospin zero and one for 4 nucleon systems: near degeneracies

Arun Kingan, and Larry Zamick

Department of Physics and Astronomy

Rutgers University, Piscataway, New Jersey 08854

## Abstract

In this work we calculate the energies odd $J^+$ states in selected even-even nuclei- $^{44}$Ti $^{52}$Fe and $^{96}$Cd. Of particular interest is the fact that in many cases the first T=0 and the first T=1 state are close in energy. Whether the lowest odd J state has T=0 or T=1 is here considered.

*Keywords:* Odd $J^+$ states,

PACs Number: 21.60.Cs

# 1 Introduction

Most attention in even-even nuclei has been given to even J states, in part because the few lowest states are indeed even J. It would be of interest now to consider odd J, even parity states. We shall do so here for systems of 2 protons and 2 neutrons in the $f_{7/2}$ shell and $g_{9/2}$ shell. These can be associated with the nuclei $^{44}$Ti ($^{52}$Fe) and $^{96}$Cd respectively. The latter is actually a 4 hole system. It is not a priori clear for these odd J states whether the lowest sate has isospin T=0 or T=1. The simplest odd-odd nucleus-the deuteron has $J=1^+$ T=0. We should add that none of the odd J states that we discuss are listed in the National Nuclear Data Center (www.nndc.org) tables, neither in ENSDF nor XUNDL. We hope this work will stimulate experimental works to fill in these surprising gaps.

# 2 Odd J Energy Levels for $^{44}$Ti $^{52}$Fe and $^{96}$Cd.

In Table I we show results of a single j shell calculation ($f_{7/2}$) of the odd $J^+$ sates in $^{44}$Ti and $^{52}$Fe using 2-body matrix elements from The spectra of $^{42}$Sc and $^{52}$Fe respectively. These were previously used for calculations in the $f_{7/2}$ shell by Zamick and Robinson [1]. The former is called the MBZE interaction and detailed wave functions are shown [2]. This has improved T=0 two-body matrix elements relative to earlier calculations [3,4].

It should be noted that in single j shell all the $J=1^+$ states have isospins T=1, likewise the maximum $J_{odd}=11^+$ state. Aside from these all the odd J states in $^{44}$Ti calculations ($J=3^+,5^+,7^+,9^+$) have T=0 as the lowest state. If we had used the same 2 body matrix elements for $^{52}$Fe we would have had identical results (hole-hole interaciton is the same particle particle). However we find the opposite –for $^{52}$Fe the $J=3^+,5^+,7^+$ and $9^+$ T=1 is the lowest.

In the full f-p calculations shown in Table 2 the same is true - T=0 is lowest for $J=3^+,5^+,7^+$ and $9^+$. A new feature is that we now have T=0 states with $J=1^+$ and $J=11^+$ which were not present in single j. These states lie considerably higher than the corresponding T=1 states.

In the $g_{9/2}$ shell, as shown in Table 3, in single j the interaction of Corragio et al. was used.[5]. We have a reversal compared to $^{44}$Ti. The $J=3^+,5^+,7^+$ and $9^+$ states with isospins T=1 lie lower that the corresponding

T=0 states. For J=11$^+$ and 13$^+$ the T=0 states are the lowest. Also of course with states of different isospin so close there could be strong Coulomb mixing effects coming into play.

In the large space for $^{96}$Cd with the jj44b interaction, Table 4, there are many reversals. For all odd J states in Table IV the T=0 states are the lowest, except for J=1$^+$ and J=9$^+$. For these two, T=1 is the lowest. For J=1$^+$ this is understandable because in the small space there is no J=1$^+$ T=0 state, so one has to go to higher configurations to get this state. For J=9$^+$ and J=11$^+$ there is a near degeneracy of the 2 states-T=0 and T=1. It would be of interest to see if there is indeed a near degenerate doublet in $^{96}$Cd with J=9$^+$ and J=11$^+$.

Details of the interactions are contained in references [6], [7], [8] and [9].

.

# 3 Magnetic Dipole Transitions

The competition for the isospin of the ground states of N=Z odd-odd nuclei has recieved considerable attention over the years. As an example the following nuclei have T=0 ground states: deuteron (J=1$^+$), $^{18}$F (J=1$^+$), $^{22}$Na (J=3$^+$) and $^{38}$K(J=3$^+$). But then there is a switch to T=1 for $^{42}$Sc, $^{46}$V, $^{50}$Mn and $^{54}$Co. All these have J=0$^+$ T=1 ground states. But to make things more complicated for $^{58}$Cu we are back to T=0 (J=3$^+$). In this work we extend the competition to higher values of J.

The sum of the proton and neutron magnetic moments is 0.88 $\mu_n$ while the difference is 4.706 $\mu_n$ [10].. This leads to the fact that in $^{44}$Ti and $^{46}$Cd B(M1)'s involving changes in isospin i.e. from T=1 to T=0 or from T=0 to T=1 are in general much larger than those from T=0 to T=0. Also in the single j shell from T=1 to T=1 in N=Z nuclei the B(M1)'s vanish. In this section, we show in tables V through XIV some sample B(M1)'s to illustrate this point. Perhaps measuring B(M1)'s will help experimentalists to determine the spin and isospins of some of the odd J states. The states have a rather simple structure so it would be a shame not to look for them.

Table 1:Excitation energies (MeV) of odd J, T=0 and T=1 states in $^{44}$Ti and $^{52}$Fe-Single j

| $^{44}$Ti | | | $^{52}$Fe | | |
|---|---|---|---|---|---|
| J | T=0 | T=1 | J | T=0 | T=1 |
| 1$^+$ | — | 5.669 | 1$^+$ | — | 5.442 |
| 3$^+$ | 5.786 | 6.001 | 3$^+$ | 6.540 | 5.834 |
| 5$^+$ | 5.871 | 6.512 | 5$^+$ | 6.601 | 6.463 |
| 7$^+$ | 6.043 | 6.501 | 7$^+$ | 6.017 | 5.890 |
| 9$^+$ | 7.984 | 8.626 | 9$^+$ | 8.047 | 7.791 |
| 11$^+$ | — | 9.806 | 11$^+$ | — | 8.666 |

Table 2:Excitation energies (MeV) of odd J, T=0 and T=1 states in $^{44}$Ti, full f-p shell calculation using fpd6

interaction.

| J | T=0 | T=1 |
|---|---|---|
| 1+ | 9.533 | 7.349 |
| 3+ | 6.118 | 6.914 |
| 5+ | 7.264 | 7.418 |
| 7+ | 7.744 | 7.338 |
| 9+ | 8.828 | 9.124 |
| 11+ | 10.815 | 10.182 |

Table 3:Excitation energies (MeV) of odd J, T=0 and T=1 states in $^{96}$Cd, single j CCGI

| J | T=0 | T=1 |
|---|---|---|
| 1+ | — | 4.269 |
| 3+ | 4.956 | 4.467 |
| 5+ | 4.936 | 4.566 |
| 7+ | 4.927 | 4.635 |
| 9+ | 4.879 | 4.365 |
| 11+ | 5.922 | 5.964 |
| 13+ | 6.107 | 6.483 |
| 15+ | — | 6.550 |

Table 4:Excitation energies (MeV) of odd J, T=0 and T=1 states in $^{96}$Cd, full shell calculation using the jj44b interaction

| J | T=0 | T=1 |
|---|---|---|
| 1+ | 5.6556 | 4.5472 |
| 3+ | 4.3338 | 4.6643 |
| 5+ | 4.8553 | 5.0529 |
| 7+ | 4.9706 | 5.199 |
| 9+ | 4.8824 | 4.777 |
| 11+ | 6.0641 | 6.1567 |
| 13+ | 6.5172 | 7.0021 |
| 15+ | — | 7.1475 |

Odd J states are of interest in a different context. In works of Moya de Guerra et al. [11], Qi et al. [12] and Zamick and Escuderos [13]. the problem of odd J pairing is addressed. The quantity $J_{max}$ is equal to 2j and is odd. e.g. $J_{max}=7$ in the $f_{7/2}$ shell and 9 in $g_{9/2}$. We define a $J_{max}$ interaction as one in which all 2 body matrix elements are zero except when $J=J_{max}$ and here we have an attractive interaction. It was shown in [11] and especially in [12] that the wave functions are often well approximated by properly normalized unitary 9j coefficients and more to the point have high overlaps with wave functions of more realistic interactions. However in was shown in [13] that the spectrum with such an interaction is very bad. In say $^{44}$Ti the ground state has the largest possible spin $J=12^+$. One gets a sort of inverted spectrum.

Table 5: $^{44}$Ti B(M1)'s from $J=1^+$ to $J=0^+$ using fpd6 and fpd6 with charge dependance

| No charge dependance | | | | Charge dependent | | |
|---|---|---|---|---|---|---|
| Isospin | 1 | 0 | – | | | |
| v Isospin | Energy | 9.2653 | 9.5326 | Energy | 9.2374 | 9.3891 |
| 0 | 0 | 0.0848 | 0.0003 | 0 | 0.0707 | 0.0019 |
| 0 | 5.2738 | 0.2459 | 0.0000 | 5.1902 | 0.2638 | 0.0001 |
| 0 | 8.4598 | 1.6700 | 0.0000 | 8.4738 | 1.5860 | 0.0061 |
| 2 | 9.2428 | 0.1288 | 0.0000 | 9.1985 | 0.0325 | 0.0073 |
| 0 | 9.7493 | 0.1328 | 0.0007 | 9.6328 | 0.1421 | 0.0019 |
| 1 | 10.3587 | 0.0001 | 0.0010 | 10.2477 | 0.0294 | 0.0067 |

Table 6: $^{44}$Ti B(M1)'s from $J=1^+$ to $J=2^+$ using fpd6 and fpd6 with charge dependence

| No charge dependance | | | | Charge dependent | | |
|---|---|---|---|---|---|---|
| Isospin | 1 | 0 | – | | | |
| Isospin | Energy | 9.2653 | 9.5326 | Energy | 9.2374 | 9.3891 |
| 0 | 1.3001 | 0.2990 | 0.0000 | 1.2678 | 0.3111 | 0.0000 |
| 0 | 4.3356 | 0.3494 | 0.0005 | 4.3392 | 0.3800 | 0.0009 |
| 1 | 6.1072 | 0.0007 | 0.4453 | 6.1299 | 0.0001 | 0.3729 |
| 0 | 7.0658 | 0.0004 | 0.0000 | 6.9528 | 0.0028 | 0.0000 |

Table 7: $^{44}$Ti B(M1)'s from $J=3^+$ to $J=2^+$ using fpd6 and fpd6 with charge dependence

| No charge dependance | | | | Charge dependent | | |
|---|---|---|---|---|---|---|
| Isospin | 0 | 1 | – | | | |
| Isospin | Energy | 6.1178 | 6.914 | Energy | 6.0724 | 6.885 |
| 0 | 1.3 | 0.0000 | 0.2276 | 1.2678 | 0.0002 | 0.2445 |
| 0 | 4.3356 | 0.0004 | 0.5706 | 4.3392 | 0.0007 | 0.5430 |
| 1 | 6.1072 | 0.0004 | 0.0004 | 6.1299 | 0.0001 | 0.0000 |
| 0 | 7.0658 | 0.0003 | 0.2518 | 6.9528 | 0.0011 | 0.2371 |

Table 8: $^{44}$Ti B(M1)'s from $J=3^+$ to $J=4^+$ using fpd6 and fpd6 with charge dependence

| No charge dependance | | | Charge dependent | | |
|---|---|---|---|---|---|
| Isospin | 0 | 1 | – | | |
| Isospin | Energy | 6.1178 | 6.914 | Energy | 6.0724 | 6.885 |
| 0 | 2.4983 | 0.0002 | 0.2716 | 2.468 | 0.0004 | 0.2718 |
| 0 | 5.4297 | 0.0004 | 0.1158 | 5.4218 | 0.0005 | 0.1086 |
| 1 | 6.6365 | 0.3783 | 0.0008 | 6.6426 | 0.3772 | 0.0321 |
| 0 | 6.7303 | 0.0001 | 1.0580 | 6.7411 | 0.0155 | 1.0020 |

Table 9: $^{44}$Ti B(M1)'s from J=5$^+$ to J=**6$^+$** using fpd6 and fpd6 with charge dependence

| No charge dependance | | | Charge dependant | | |
|---|---|---|---|---|---|
| Isospin | 1 | 0 | – | | |
| Isospin | Energy | 7.2636 | 7.418 | Energy | 7.2201 | 7.3979 |
| 0 | 3.7756 | 0.4021 | 0.0005 | 3.7727 | 0.4049 | 0.0008 |
| 0 | 6.6186 | 0.7243 | 0.0011 | 6.627 | 0.6958 | 0.0012 |
| 1 | 6.7579 | 0.0003 | 0.2206 | 6.774 | 0.0006 | 0.2694 |
| 0 | 7.1017 | 0.1648 | 0.0009 | 7.0987 | 0.1643 | 0.0017 |

Table 10: $^{44}$Ti B(M1)'s from J=7$^+$ to J=6$^+$ using fpd6 and fpd6 with charge dependence

| No charge dependance | | | Charge dependant | | |
|---|---|---|---|---|---|
| Isospin | 0 | 1 | – | | |
| Isospin | Energy | 7.3376 | 7.7437 | Energy | 7.3002 | 7.7499 |
| 0 | 3.7756 | 0.4768 | 0.0000 | 3.7727 | 0.4741 | 0.0003 |
| 0 | 6.6186 | 0.4496 | 0.0000 | 6.627 | 0.4526 | 0.0003 |
| 1 | 6.7579 | 0.0007 | 0.3513 | 6.774 | 0.0186 | 0.3627 |
| 0 | 7.1017 | 0.8997 | 0.0003 | 7.0987 | 0.8717 | 0.0058 |

Table 11: $^{44}$Ti B(M1)'s from J=11$^+$ to J=**10$^+$** using fpd6 and fpd6 with charge dependence

| No charge dependance | | | Charge dependant | | |
|---|---|---|---|---|---|
| Isospin | 1 | 0 | – | | |
| Isospin | Energy | 10.1818 | 10.815 | Energy | 10.1796 | 10.7181 |
| 0 | 7.6135 | 0.0921 | 0.0003 | 7.6051 | 0.0907 | 0.0000 |
| 0 | 9.9291 | 3.4540 | 0.0001 | 9.9164 | 3.3950 | 0.0032 |
| 1 | 10.5459 | 0.0001 | 0.1382 | 10.5517 | 0.0216 | 0.1568 |
| 0 | 10.6341 | 0.2929 | 0.0019 | 10.5696 | 0.2909 | 0.0002 |

Table 12: $^{44}$Ti B(M1)'s from J=11$^+$ to J=12$^+$ using fpd6 and fpd6 with charge dependence

| No charge dependance | | | Charge dependant | | |
|---|---|---|---|---|---|
| Isospin | 1 | 0 | – | | |
| Isospin | Energy | 10.1818 | 10.815 | Energy | 10.1796 | 10.7181 |
| 0 | 8.312 | 3.2960 | 0.0005 | 8.3174 | 3.2960 | 0.0058 |
| 0 | 13.0119 | 0.1049 | 0.0000 | 13.0346 | 0.1031 | 0.0003 |
| 1 | 14.7409 | 0.0014 | 0.0468 | 14.7538 | 0.0025 | 0.0432 |
| 0 | 18.5681 | 0.0156 | 0.0000 | 18.5921 | 0.0155 | 0.0000 |

As shown in Tables XI and XI, although there are quantitative differences between the B(M1)'s from J=11$^+$ to J=10$^+$ using fpd6 with and without charge dependance, there are no qualitative differences. Despite the arguement that the large diffeence in isoscalar and isovector magnetic coupling the B(M1)'s with and without charge independence are nearly the same. We here offer a rough argument. In single j shell the main effect of the Coulomb interaciton is to provide a repulsive term to the J=0 2 body matrix element–a sort of antipairing in the proton-proton channel. Fot odd j staes in say $^{44}$Ti there are no amplitudes where hte 2 protons couple to zero. This is because if indeedin the satee $[J_P J_n\}J$ we have $J_p=0$, the J must be $J_N$. But $J_N$ for 2 particles can only be even. Thus the odd J states miss the main effect of the Coulomb interaction. There also no $J_P = 0$ components for even J states in $^{44}$Ti with J greater than 6. So they will also not feel the brunt of the Coulomb interaction. The most promising case in $^{44}$Ti or $^{52}$Fe is J=6 where the T=1 –T==0 spliting is smallest ( 5.612 -4.012= 1.6 MeV).

For other even J values the splittings are greater.. In the NNDC tables the lowest 2$^+$ T=0 state is at 1.087 MeV and the lowest 2$^+$ T=1 state is at 6.606 MeV. With fpd6 the respective numbers are 1.300 Mev and 6.107 MeV. The lowness of the T=1 state may be due to the fact that there are deformed admixtures in the ground state of $^{44}$Ti. This nucleus is not included in the fit of M. Homma et al. [9] Evidence of these deformed admixtures is shown in many references but we only include one of the latest by K. Arnswald et al. [13]. A most promising case for large

One can get good estimates of higher isospin state energies using formulas involving binding energies and Coulomb energies, e.g. as shown by Zamick et al. [10]

Table 13: $^{96}$Cd B(M1)'s from J=9$^+$ to J=8$^+$ using jj44b interaction

| Isospin | 1 | 0 | |
|---|---|---|---|
| Isospin | Energy | 4.777 | 4.8824 |
| 0 | 3.4047 | 0.1857 | 0.00003978 |
| 1 | 4.1928 | 0.00000234 | 0.9273 |
| 0 | 4.4348 | 0.3345 | 0.00001192 |
| 0 | 5.1582 | 1.507 | 0.000519 |
| 0 | 5.6245 | 0.585 | 0.00003159 |
| 1 | 5.8083 | 9.089E-07 | 1.644 |

Table 14: $^{96}$Cd B(M1)'s from J=9$^+$ to J=10$^+$ using jj44b interaction

| Isospin | 1 | | 0 | |
|---|---|---|---|---|
| Isospin | Energy | 4.777 | | 4.8824 |
| 0 | 4.7891 | 1.49 | | 7.063E-07 |
| 0 | 5.3702 | 0.03691 | | 0.00003068 |
| 0 | 5.628 | 1.249 | | 0.00002998 |
| 1 | 6.024 | 0.00000591 | | 1.858 |
| 0 | 6.4612 | 1.328 | | 0.00004016 |
| 0 | 6.824 | 0.07812 | | 4.326E-06 |
| 1 | 6.8572 | 7.515E-07 | | 0.103 |

# 4 $^{48}$Cr levels in single j

Single j shell calculations (j=$f_{7/2}$) for levels of $^{48}$Cr having presented in the archives by Escuderos et al. [2]. We present in Table XV the energies of the first 5 states of each odd J.

T=1 states are designated with a * symbol. The others have T=0. We note that the states of J=3$^+$ to 13$^+$ have isospin T=0 for the lowest state. For the extreme spins, J=1$^+$ and 15$^+$ the lowest states have isospins T=1. It is interesting to note that $^{48}$Cr is the lightest even-even nucleus to have J=1$^+$ T=0 states in a single j shell configuration. There are no T=0 J=1$^+$ states of the $(f_{7/2})^4$ configuration in $^{44}$Ti. We note that for odd J there tends to be a pile up of states.

Table 15: T=0 and T=1 ( *) states in $^{48}$Cr with the MBZE interaction–single j configuration.

| J | ENERGY (MeV) | | | | |
|---|---|---|---|---|---|
| 1 | 5.479* | 7.015* | 7.484* | 7.775 | 8.244 |
| 3 | 5.624 | 5.919 | 5.954* | 6.083* | 6.434 |
| 5 | 4.295 | 5.405 | 5.789* | 6.119* | 6.461 |
| 7 | 5.955 | 6.329 | 6.359* | 7.922 | 7.952 |
| 9 | 6.989 | 7.441 | 7.857* | 8.491 | 9.228* |
| 11 | 8.619 | 9.612* | 10.426 | 10.537* | 11.168 |
| 13 | 11.567 | 11.888 | 11.924* | 13.183* | 14.528* |
| 15 | 14.539* | | | | |

.

# 5 Sensitivity to the interactions

.

. In this section we wish to show that getting a simple result, i.e near degeneracy of T=0 and T=1 odd J states, does not have a simpe explanaiton. One often talks about T=1 vs. T=0 pairing. In Table 16 we show the energysplitting E(T=0) -E(T=1) for various scematic interactions and compare with the interaction of ref [2] where matrix elements were fit to experiment. The 8 interaction matrix elements for the $(f_{7/2})^2$ configuration corresponding to J=0,1,2,3,4,5,6,7 are MeV)

INTA : (J=0,T=1 ) -1,0,0,0,0,0,0,0

INTB :($J_{max}$ T=0) 0,0,0,0,0,0,0,-1

INTC :( Q.Q) 0, 0.4096, 1.1471 , 2.0483, 2.8677, 3.2774, 2.8677, 1.1471

INTD:( MBZE) 0, 0.6110,1.5863, 1.4904, 2.8153, 1.5101,3.2420, 0.6163

.

.

Table 16 : E(T=0)-E(T=1) for various interactions.

| J | J=0 T=1 | $J_{max}$ T=0 | Q.Q | MBZE |
|---|---|---|---|---|
| 3 | 0.75 | 0.0467 | 1.9811 | -0.2151 |
| 5 | 0.75 | 0.0299 | 0.6223 | -0.6409 |
| 7 | 0.75 | 0.0023 | 0.3346 | -0.4661 |
| 9 | 0 | 0.0511 | 1.6387 | -0.6425 |

+

.The strengh of Q.Q is such that the lowest $2^+$ state is at 0.9665 MeV.For Q.Q and MBZE a constant was added so as to have the $J=0^+$ state come at zero energy. This does not affect the energy differences.

We get quite a variety of answers -significanly positive for (J=0 T=1) and QQ, vey small but positive for ($J_{max}$ T=0) but negative for the realistic MBZE .This tells us that the energy difference tests the fine tuning of the interaction that is used. One cannot simply wave one's hands and talk about T=1 pairing or T=0 pairing. All detais of the interaction come into play.One cannot use binding energy arguments to understand the energy differences that we are dealing with here. Biding enrgies involve verages over 2 body matrix elements. Here we are dealing with fluctuations awaty from the average.

Note that one gets positive result for both (J=0 T=1) and ($J_{max}$ T=0).They don't cancel each other out. The results for (J=0 T=1) come from the analytic formula of Edmunds and Flowers [15,16],and is used also in works of Neergaard[17] and Harper and Zamick[18].

# 6 Previous results on isobaric analog states in odd-odd nuclei

As a counterpoint we here discuss a related problem where binding energies comevery much into play. In a previous work [19] we considered the energies of isobaric analog states in odd-odd nuclei. In that work there were shell model estimats and mass formula estimates. We briefly show in Table 17 some selected results. Beyond the early work of Bethe and Bacher[20] the elaborate mass formula of Dulco and Zucker (D-K) was used[21]. The results are in the right ball park but there are significant deviations . In all but [52]Mn the shell model resuits give energies that are too high whereas D-K binding energy gives results that are too low for all nuclei here considered. We find it useful to show Table 17 just to show what is now the state of the art. It

Table 17 Energies of Isoaric Analog States in Odd-Odd Nuclei (MeV)

| Nucleus | Shell Model | D-K | Experiment |
|---|---|---|---|
| $^{44}$Sc | 3.418 | 2.351 | 2.719 |
| $^{46}$Sc | 5.230 | 4.694 | 5.022 |
| $^{52}$Mn | 2.731 | 1.881 | 2.926 |
| $^{60}$Cu | 2.726 | 2.347 | 2.530 |

# 7 Additional Remarks

.We have here pointed out that there are very simple stuctures which for the most part have not yet been found in experiment -Odd J states. One reason is that they lie at higher energies than the lowest few even J states. Because T=0 and T=1 odd J states often have small energy splittings they are good candidates for studying Coulomb and other charge dependent mixings. We have presented argumets for why the mixings might be small but there is nothing like experiment to settle the issue.We emphasize that the enrgy differences cannot be obtained by mean field arguments. It is the deviations from the mean that are. important. Also we have shown that one needs the deatails of all the 2 body matrix elements in order to obtain credible values for these energy splittings.Simple arguments in terms of T=1 and/or T=0 pairing are not sufficient. But basically what we are saying is that there are simple stuctures out there that deserve further attention.

# 8 Acknowledgements

A.K. received support from the Rutgers Richard J. Plano Research Award Summer 2017. We thank H. Grawe for bringing our attention to this problem. We thank Kai Neergï¿œrd for valuable discussions.

# References

[1] L.Zamick and S.J.Q. Robinson ,Physics of Atomic Nuclei 65, 054321 (2002)

[2] A. Escuderos, L.Zamick and B.F. Bayman, Wave functions in the $f_{7/2}$ shell, for educational purposes and ideas, arXiv: nucl-th/ 05060

[3] J.D. McCullen, B.F. Bayman, and L. Zamick, Phys Rev **134**, B515 (1964)

[4] J.N. Ginocchio and J.B. French , Phys. Lett **7**, 137 (1963)

[5] L. Coraggio, A. Covello, A. Gargano, and N. Itaco, Phys. Rev. C 85, 034335 (2012).

[6] W.A. Ormand and B.A. Brown, Nucl. Phys.A 491,1(1989)

[7] B.A. Brown and W.D.M. Rae,The Shell-Model code NUSHELLX@MSU, http://www.sciencedirect.com/science/article/pii/S0090375214004748

[8] W.A.Richter ,M.G. Van Der Merwe, R.E. Julies and B.A. Brown, Nucl. Phys. A523, 325 (1991)

[9] M. Homma, T Otsuka, B. A. Brown and T. Mizusaki PRC 69, 034335 (2004)

[10] L. Zamick. Y.Y. Sharon and S.J.Q. Robinson, Phys. Rev. C82, 087303 (2010)

[11] E. Moya de Guerra, A.A. Raduta, L. Zamick, and P. Sarriguren, Nuclear Physics A 727, 3

[12] C. Qi, J. Blomqvist, T. Bï¿œck, B. Cederwall, A. Johnson, R.J. Liotta, and R. Wyss, Phys. Rev. C 84, 021301(R) (2011).

[13] L.Zamick and A. Escuderos ,Phys. Rev. C 87, 044302 (2013)

[14] K Arnswald et al. Phys Lett B **772**, 599 (2017)

[15] B.H Flowers, Proc. Roy. Soc. (London) A212, 248 (1952)

[16] A.R. Edmonds and B. H. Flowers, Proc. Roy. Soc. (London) A214, 515 (1952)

[17] K. Neergaard,Phys. Rev. C90,0143018 (2014)

[18] M. Harper and L.Zamick Phys. Rev. C91,014304 (2015)

[19] L.Zamick, S.J.Q.Robinson., A. Escuderos, Y.Y. Sharon and M. Kirson ,Phys. Rev. C 87, 067302 (2013)

[20] H.A. Bethe and B.F. Bacher, Rev. Mod. Phys.82,81(1936)

[21] J.Duflo and A.P.Zuker, Phys. Rev. C52, R23 (1995)

[22] G. Audi, F.G.Kondev, M.Wang, B.Pfeiffer, X.Sun, J.Blanchot,and M. MacCormick, Chinese Physics (HEP and NP), 36(12), 1157 (2012)